\documentclass[aps,prl,superscriptaddress,twocolumn,showpacs]{revtex4}
\usepackage{amsmath,amsfonts,amssymb,bm,graphicx}
\usepackage{amsmath,amsfonts,bm,graphicx,verbatim,mathrsfs,units}
\usepackage[colorlinks=true,citecolor=blue]{hyperref}

\begin{document}

\title{Full Counting Statistics of Andreev Tunneling}
\author{Ville F.~Maisi}
\affiliation{Low Temperature Laboratory (OVLL), Aalto University School of Science, P.O. Box 13500, 00076 Aalto, Finland}
\affiliation{Centre for Metrology and Accreditation (MIKES), P.O. Box 9, 02151 Espoo, Finland}
\author{Dania Kambly}
\affiliation{D\'epartement de Physique Th\'eorique, Universit\'e de Gen\`eve, 1211 Gen\`eve, Switzerland}
\author{Christian Flindt}
\affiliation{D\'epartement de Physique Th\'eorique, Universit\'e de Gen\`eve, 1211 Gen\`eve, Switzerland}
\author{Jukka P.~Pekola}
\affiliation{Low Temperature Laboratory (OVLL), Aalto University School of Science, P.O. Box 13500, 00076 Aalto, Finland}


%72.70.+m 	Noise processes and phenomena
%73.23.-b 	Electronic transport in mesoscopic systems
%73.23.Hk 	Coulomb blockade; single-electron tunneling
%74.50.+r 	Tunneling phenomena; Josephson effects

\newcommand{\llangle}{\langle\!\langle}
\newcommand{\rrangle}{\rangle\!\rangle}

\begin{abstract}
We employ a single-charge counting technique to measure the full counting statistics (FCS) of Andreev events in which Cooper pairs are either produced from electrons that are reflected as holes at a superconductor/normal-metal interface or annihilated in the reverse process. The FCS consists of quiet periods with no Andreev processes, interrupted by the tunneling of a single electron that triggers an avalanche of Andreev events giving rise to strongly super-Poissonian distributions.
\end{abstract}

\date{\today}
\maketitle

Superconductors are materials that below a critical temperature lose their electrical resistance and thereby allow a supercurrent to flow \cite{Tinkham2004}. Inside the superconducting gap electrons combine into Cooper pairs that carry electrical charge through the superconductor without dissipation.  The conversion of a Cooper pair into normal-state electrons (or vice versa) is known as an Andreev process \cite{Andreev1964}. In a direct Andreev process, an electron in a normal-state material is reflected as a hole at the interface with a superconductor where a Cooper pair is formed. Moreover, with several normal-state electrodes coupled to the same superconductor, crossed Andreev reflections may occur where electrons coming from different electrodes combine into a Cooper pair.

Cooper pairs consist of highly quantum-correlated electrons and may thus serve as a source of entanglement when split into different normal-state electrodes \cite{Hofstetter2009,Hermann2010,Das2012}. The entanglement of the spatially separated electrons can be detected through current noise measurements \cite{Das2012}. Experiments on superconductor/normal-metal junctions have also revealed a doubling of the shot noise due to the conversion of Cooper pairs into normal-state electrons~\cite{Jehl2000}. However, a complete understanding of the fundamental tunneling processes at a superconductor/normal-metal interface requires measurements beyond the average current and the noise only. Higher-order correlation functions are encoded in the full counting statistics (FCS) which quantifies the probability $p(n,t)$ of observing $n$ charge transfer events during the time span $[0,t]$. The FCS of normal-state electrons has been addressed both theoretically \cite{Levitov1993,Blanter2000,Nazarov2003} and experimentally \cite{Reulet2003,Bomze2005,Gustavsson2006,Fujisawa2006,Timofeev2007,Sukhorukov2007,Gershon2008,Flindt2009,Gabelli2009,Gustavsson2009,Masne2009,Ubbelohde2012}.
In contrast, measurements of the FCS of charge transfer into superconductors have so far been lacking despite great theoretical interest \cite{Belzig2001,Belzig2001b,Borlin2002,Johansson2003,Cuevas2003,Belzig2003,Pilgram2005,Morten2008,Duarte-Filho2009,Braggio2011}.

\begin{figure}[h!]
\begin{center}
\includegraphics[width=0.8\linewidth]{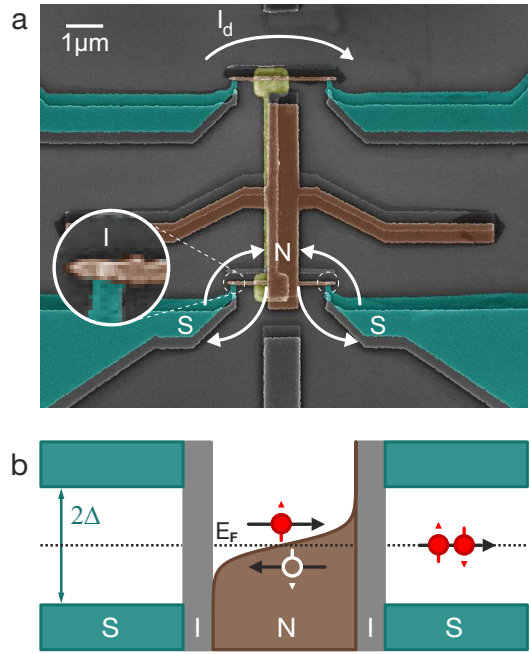}
\caption{(color online). SINIS structure and Andreev processes. {\bf a}, A metallic normal-state (N) island (brown) is connected by insulating (I) tunneling barriers to superconducting (S) leads (green). The current $I_d$ through a separate single-electron transistor (SET) is sensitive to the charge occupation of the island and is used to read out the number $N$ of excess charges on the island. A copper electrode (yellow) increases the capacitive coupling of the normal-state island to the SET and improves the detector signal-to-noise ratio.  {\bf b}, An electron above the Fermi level of the normal-state island is reflected as a hole and a Cooper pair is formed in one of the superconductors. Without a voltage across the SINIS, the Fermi energy $E_F$ of the normal-state material lies in the middle of the superconducting gap $2\Delta$.} \label{Fig1}
\end{center}
\end{figure}

\begin{figure*}
\begin{center}
\includegraphics[width=0.95\linewidth]{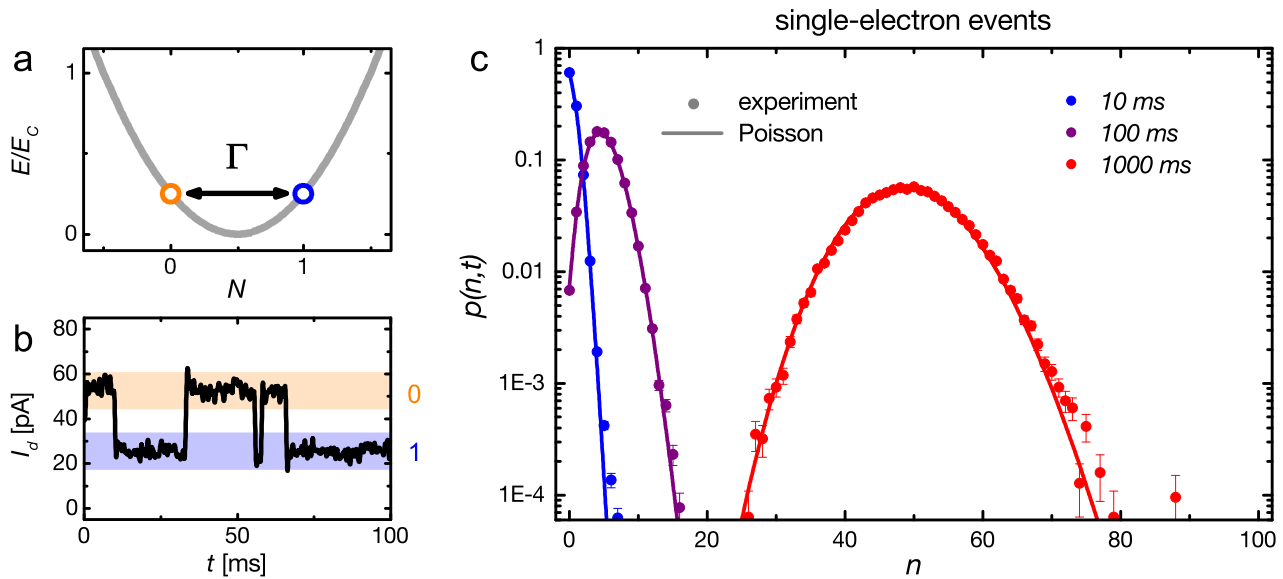}
\caption{(color online). FCS of single-electron events. {\bf a}, Charging diagram showing Eq.~(\ref{eq:charging_energy}) with $N_g = 0.5$. The charge states with $N=0$ or $N=1$ excess charges on the island are degenerate. The transitions $0 \leftrightarrows 1$ occur with rate $\Gamma = 49\ \mathrm{Hz}$. Other charge states are energetically unfavorable. {\bf b}, Time trace of the current $I_d$ in the SET-detector, which switches between two levels corresponding to $N=0$ and $N=1$, respectively. {\bf c}, Measured FCS of single-electron events for different observation times $t= 10$, 100, and 1000 ms. Poisson distributions given by Eq.~(\ref{eq:poisson}) are shown with full lines.} \label{Fig2}
\end{center}
\end{figure*}

In this Letter we report measurements of the FCS of Andreev events occurring between a normal-metal island and two super-conducting leads. Our measurements of the FCS allow us to develop a detailed understanding of the elementary tunneling processes at the superconductor/normal-metal interfaces. Figure \ref{Fig1}a shows our SINIS structure consisting of a normal-state copper island~(N) connected by insulating~(I) aluminum-oxide tunnel barriers (of a few nanometers thickness \cite{Prunnila2010}) to a pair of superconducting~(S) aluminum leads. The structure was patterned on an oxidized silicon chip using standard $e$-beam lithography techniques. A copper coupling strip was first formed and covered with a 50 nm thick aluminum-oxide layer grown by atomic layer deposition. Gold leads (not shown) were then patterned, making a direct metallic contact to the superconducting leads. Finally, the SINIS structure and the gate leads were formed by $e$-beam evaporation at different angles. Tunnel barriers were created by thermal oxidation in between.

The number of excess electrons $N$ on the island is discrete and can be controlled by applying a voltage $V_g$ to a gate electrode below it. We parameterize the off-set voltage by the variable $N_g=C_g V_g/e$, where $C_g$ is the gate capacitance and $e$ the electronic charge. The energy required for charging the island with $N$ electrons is~\cite{Nazarov2009}
\begin{equation}
E = E_c (N-N_g)^2,
\label{eq:charging_energy}
\end{equation}
where the charging energy $E_c=e^2/2C_{\Sigma}$ contains the total island capacitance $C_{\Sigma}$. The structure was designed to have a large capacitance, such that the charging energy is smaller than the superconducting gap $\Delta$ of aluminum, thereby allowing for Andreev processes to occur between the island and the superconducting leads, Fig.~\ref{Fig1}b. The charging energy $E_c = 40\ \mathrm{\mu eV}$, the superconducting gap $\Delta = 210\ \mathrm{\mu eV}$ and the tunnel resistance $R_T = 490\ \mathrm{k \Omega}$ were determined by measuring the current-voltage characteristics of the SINIS structure.  Measurements were performed in a dilution refrigerator at $50\ \mathrm{mK}$ bath temperature. The charge state of the island was monitored using a nearby single-electron transistor (SET), whose conductance depends strongly on the number of excess charges on the island \cite{Gustavsson2006,Fujisawa2006,Sukhorukov2007,Flindt2009,Gustavsson2009,Maisi2011,Ubbelohde2012,House2013}.

To illustrate the basic operating principle of our device we first tuned the off-set voltage to $N_g = 0.5$. Figure~\ref{Fig2}a shows the energy for different numbers of excess charges. The states $N = 0$ and $N = 1$ are degenerate, while all other charge states are energetically unfavorable. In this case, single electrons may tunnel on and off the island from the aluminum leads with rate $\Gamma$. The origin of the single-electron tunneling is addressed in Ref.~\cite{Saira2012}. Figure~\ref{Fig2}b shows a measured time trace of the current  $I_d$ in the SET-detector, which switches between two values corresponding to $N = 0$ and $N = 1$. We count the number of single-electron tunneling events on and off the island. No voltage bias is applied. Figure~\ref{Fig2}c displays the measured distribution $p(n,t)$ of the number $n$ of single-electron events that have occurred during the time span $[0,t]$. The mean number of events increases with the observation time $t$ and the distribution grows wider. The single-electron events are uncorrelated and should be distributed according to a Poisson distribution
\begin{equation}
p(n,t)=\frac{(\Gamma t)^n}{n!}e^{-\Gamma t}
\label{eq:poisson}
\end{equation}
with mean $\langle n\rangle=\Gamma t$. From this mean value we can extract the tunneling rate $\Gamma$. Figure~\ref{Fig2}c then shows that the FCS of single-electron events indeed is well-captured by the Poisson distribution above.

\begin{figure*}
\begin{center}
\includegraphics[width=0.95\linewidth]{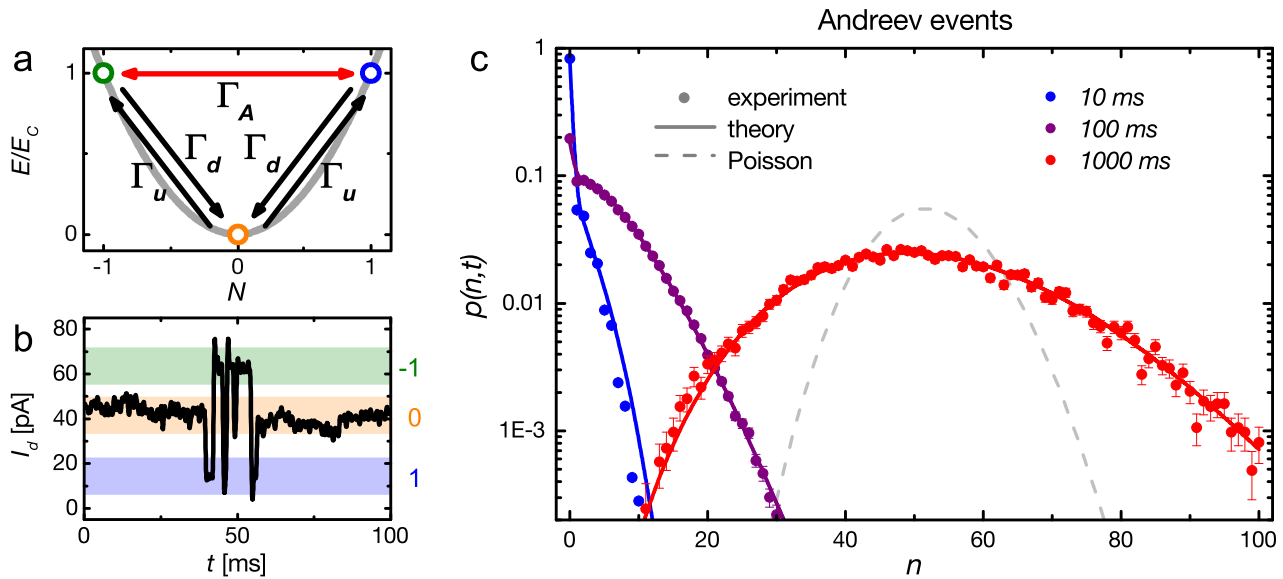}
\caption{(color online). FCS of Andreev events.  {\bf a}, Charging diagram showing Eq.~(\ref{eq:charging_energy}) with $N_g = 0$. In the ground state, the island is occupied by $N=0$ excess charges. A single-electron event may bring the island to a state with $N=\pm 1$ excess charges. The excitations $0 \rightarrow \pm 1$ occur with the rate $\Gamma_u = 12\ \mathrm{Hz}$. Relaxation to the ground state $\pm 1 \rightarrow 0$ happens with the rate $\Gamma_d=252\ \mathrm{Hz}$. The transitions $-1 \leftrightarrows 1$ correspond to Andreev events with the rate $\Gamma_A=615\ \mathrm{Hz}$. {\bf b}, Time trace of the  SET-detector current $I_d$, which switches between three levels corresponding to $N=-1$, $N=0$, and $N=1$, respectively. {\bf c}, Measured FCS of Andreev events for different observation times $t= 10$, 100, and 1000 ms.  Full lines are calculations based on Eqs.~(\ref{eq:master_eq},\ref{eq:rate_mat}). For comparison a Poisson distribution corresponding to 1000 ms is shown with a dashed line.}  \label{Fig3}
\end{center}
\end{figure*}

We are now ready to measure the FCS of Andreev events. To this end, we tuned the off-set voltage to $N_g = 0$. In this case, the charging diagram in Fig.~\ref{Fig3}a is slightly more involved: The lowest-energy state of the system is the configuration with $N=0$ excess charges. However, a single-electron event may bring the system to one of the excited states with $N=\pm1$ excess charges. The excited states are energetically degenerate and the island can make transitions between $N=-1$ and $N=1$ through Andreev processes, where  \emph{two} electrons at a time are converted into a Cooper pair in one of the superconductors or vice versa. The Andreev events occur with an average rate $\Gamma_A$ until the system relaxes back to the ground state through a single-electron event. The current $I_d$ in the SET-detector now switches between three different values corresponding to $N=-1$, 0, or 1, see Fig.~\ref{Fig3}b. (A fast sequence of single-electron events, $-1\rightarrow 0\rightarrow 1$, may be mistaken for an Andreev process, $-1\rightarrow 1$, although it is unlikely.) We count the number of Andreev tunneling events to and from the island.  Figure~\ref{Fig3}c shows the measured FCS of Andreev events obtained from around $640~000$ Andreev processes. Again, the mean value of Andreev events grows with time, however, compared to the FCS of single-electron events, the width of the distributions is surprisingly large and the FCS is strongly super-Poissonian.

To understand quantitatively the FCS of Andreev events, we consider the probabilities $p_0(n,t)$ and $p_A(n,t)$ of the island being in the ground state or in one of the excited states, where Andreev events are possible. Both probabilities are resolved with respect to the number $n$ of Andreev events that have occurred during the time span $[0,t]$. The FCS of Andreev events is $p(n,t)= p_0(n,t)+p_A(n,t)$, which can be conveniently expressed as the inner product $p(n,t)= \langle \tilde{0}|p(n,t)\rangle$ of the vectors $\langle \tilde{0}|=[1,1]$ and $|p(n,t)\rangle = [p_A(n,t),p_0(n,t)]^T$ \cite{Flindt2004,Jauho2005}. We also introduce the moment generating function $\mathcal{M}(\chi,t)=\sum_{n=0}^\infty p(n,t)e^{in\chi}=\langle\tilde{0}|p(\chi,t)\rangle$ with $|p(\chi,t)\rangle=\sum_{n=0}^\infty e^{in\chi}|p(n,t)\rangle$. The master equation for $|p(\chi,t)\rangle$ reads
\begin{equation}
\frac{d}{dt}|p(\chi,t)\rangle = \mathbb{M}(\chi)|p(\chi,t)\rangle,
\label{eq:master_eq}
\end{equation}
with the rate matrix (see also Ref.~\cite{Jordan2004})
\begin{equation}
\mathbb{M}(\chi)=\left[
                \begin{array}{cc}
                  \mathcal{H}_A(\chi)-\Gamma_d & 2\Gamma_u \\
                  \Gamma_d & -2\Gamma_u \\
                \end{array}
              \right].
\label{eq:rate_mat}
\end{equation}
Here $\mathcal{H}_A(\chi)=\Gamma_A(e^{i\chi}-1)$ is the generator of uncorrelated Andreev events occuring in the excited states with rate $\Gamma_A$. The rate for exciting the system is  $2\Gamma_u$ and $\Gamma_d$ is the relaxation rate back to the ground state, see Fig.~\ref{Fig3}a. The tunneling rates are extracted from the time traces of the SET-detector current \cite{Gustavsson2006,Fujisawa2006,Sukhorukov2007,Flindt2009,Gustavsson2009,Maisi2011,Ubbelohde2012,House2013}.
Solving Eq.~(\ref{eq:master_eq}), we find $|p(\chi,t)\rangle=e^{\mathbb{M}(\chi)t}|0\rangle$, where $|0\rangle=[2\Gamma_u,\Gamma_d]^T/(2\Gamma_u+\Gamma_d)$ is the stationary probability vector defined by $\mathbb{M}(0)|0\rangle=0$ and $\langle\tilde{0}|0\rangle=1$. The moment generating function is then $\mathcal{M}(\chi,t)=\langle \tilde{0}|e^{\mathbb{M}(\chi)t}|0\rangle$. Finally, by inverting the moment generating function for $p(n,t)$  we can evaluate the FCS of Andreev events for different observation times~$t$.

\begin{figure}
\begin{center}
\includegraphics[width=0.95\linewidth, trim = 0 0 0 0, clip]{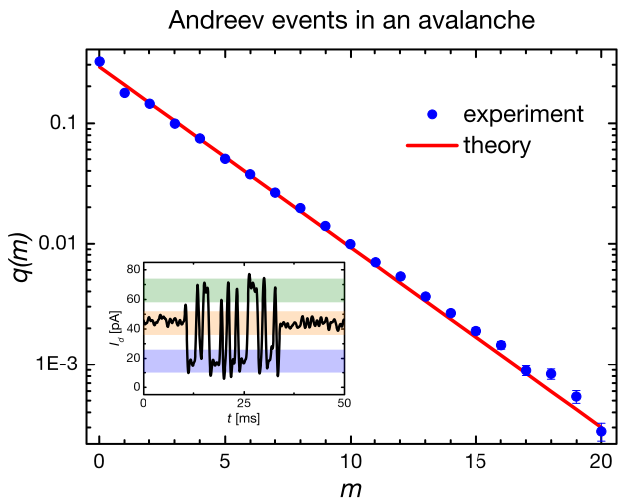}
\caption{ (color online). Number of Andreev events per avalanche. The full line indicates the theoretical prediction given by Eq.~(\ref{eq:AndreevProb}) using $\Gamma_d=252\ \mathrm{Hz}$ and $\Gamma_A=615\ \mathrm{Hz}$. The inset shows a time trace of the SET-detector current $I_d$ during an avalanche with $m=16$ Andreev events.} \label{Fig4}
\end{center}
\end{figure}

The theoretical predictions agree well with the measurements in Fig.~\ref{Fig3}c using no fitting parameters. Moreover, a physical interpretation of the non-trivial FCS follows from an expansion of the cumulant generating function $\mathcal{S}(\chi,t)=\log\{\mathcal{M}(\chi,t)\}$ in the smallest tunneling rate $\Gamma_u\ll\Gamma_d,\Gamma_A$. At long times, the cumulant generating function is determined by the eigenvalue of $\mathbb{M}(\chi)$ with the largest real-part \cite{Bagrets2003,Flindt2008}.  Importantly, the cumulant generating function for independent processes is the sum of the cumulant generating functions for the individual processes. To lowest order in $\Gamma_u$, we find at long times
\begin{equation}
\mathcal{S}(\chi,t)= 2\Gamma_u t\sum_{m=1}^{\infty}q(m)(e^{im\chi}-1)+\mathcal{O}(\Gamma_u^2)
\label{eq:CGF}
\end{equation}
with
\begin{equation}
q(m)=\frac{\Gamma_d}{\Gamma_A+\Gamma_d}\left(\frac{\Gamma_A}{\Gamma_A+\Gamma_d}\right)^m.
\label{eq:AndreevProb}
\end{equation}
This shows that the FCS can be approximated as a sum of independent Poisson processes that with rate $2\Gamma_u$ generate \emph{avalanches} of $m$ Andreev events. Each Poisson process is weighted by the probability $q(m)$ of observing an avalanche with $m$ Andreev events. In this approximation, correlations between subsequent avalanches are neglected together with the duration of the individual avalanches. These correlations would enter in Eq.~(\ref{eq:CGF}) as higher-order terms in $\Gamma_u$, but would not affect the probabilities in Eq.~(\ref{eq:AndreevProb}).  We note that similar single-electron avalanches have been predicted in molecular quantum transport \cite{Belzig2005}.

To corroborate this physical picture, we turn to the number of Andreev events per avalanche. Figure~\ref{Fig4} shows experimental results for the statistics of Andreev events within a single avalanche. The figure illustrates that avalanches with more than 10 consecutive Andreev events are possible. This is also evident from the inset showing a time trace of the detector current $I_d$ which switches 16 times between the two levels corresponding to $N=-1$ and $N=1$ excess charges, respectively. The agreement between the experimental results and the probabilities $q(m)$ in Eq.~(\ref{eq:AndreevProb}) supports the interpretation that avalanches of Andreev events, triggered by the tunneling of single electrons, give rise to the strongly super-Poissonian FCS.

In summary, we have measured the FCS of Andreev events in an SINIS structure which exhibits super-Poissonian distributions due to avalanches triggered by individual single-electron tunneling events. Our experiment opens a number of directions for future research on charge fluctuations in superconductors. These include experimental investigations of the statistics of entangled electron pairs produced in crossed Andreev reflections as well as controllable Cooper pair production and detection for quantum metrological purposes~\cite{Pekola2012}.

\emph{Acknowledgments.---} We thank O.~P.~Saira who collaborated at an early stage of the work as well as C.~Bergenfeldt and M.~B\"uttiker for instructive discussions. We acknowledge the provision of facilities and technical support by Micronova Nanofabrication Centre at Aalto University. The work was supported by the Academy of Finland through its LTQ (project no. 250280) COE grant (VFM and JPP), the National Doctoral Programme in Nanoscience, NGS-NANO (VFM) and by the Swiss NSF (DK and CF).

%\bibliography{andreevBib}

\end{document}